\documentstyle{article}
\begin{document}

\def\affil#1{\vspace*{-2.5ex}{\topsep\z@\center#1\endcenter}}
\def\altaffilmark#1{$^{#1}$}
\def\altaffiltext#1#2{\footnotetext[#1]{#2}\stepcounter{footnote}}

\def\slash#1{\setbox0=\hbox{$#1$}#1\hskip-\wd0\hbox to\wd0{\hss\sl/\/\hss}}
\def\ltsim{\mathrel{\hbox{\rlap{\hbox{\lower4pt\hbox{$\sim$}}}\hbox{$<$}}}}
\def\gtsim{\mathrel{\hbox{\rlap{\hbox{\lower4pt\hbox{$\sim$}}}\hbox{$>$}}}}

\begin{center}
{\Large Fitting Direct Interaction Pieces into Neutrino Puzzles} 
\end{center}

\begin{center}
{\large Loretta M. Johnson}\altaffilmark{1} and 
{\large Douglas W. McKay}\altaffilmark{2} \\
\em{Department of Physics and Astronomy, University of Kansas, 
Lawrence, KS 66045, USA}
\end{center}
\altaffiltext{1}{Email address: johnson@kuphsx.phsx.ukans.edu}
\altaffiltext{2}{Email address: mckay@kuphsx.phsx.ukans.edu}

\bigskip 

\noindent{\bf Abstract}

\bigskip

We develop a framework and unified, compact notation to include neutrino direct 
interaction effects with neutrino oscillations for 
a wide class of low--energy, effective four-Fermi interactions. Modest 
flavor--violating interaction
strengths can make significant changes in the boundaries in $\Delta M^2 
\leftrightarrow sin^2 2\theta$ and in $tan^2\theta_{13}\leftrightarrow
tan^2\theta_{23}$ parameter space.  
We draw examples from the $L-R$ symmetric and $MSSM + \slash R$ models and 
find the recently reported decay-at-rest 
$\mu\leftrightarrow e$ transition probability can be described without
oscillations in $L-R$ symmetric models. 

\bigskip


\section{Introduction} 
Reconciling the solar \cite{btc}, atmospheric\cite{KF}, and 
accelerator\cite{Lc,nx,kc} neutrino data is difficult within a purely 
3-family oscillation description, which  provides for two
mass-squared differences, 3 mixing angles and a CP  violating phase
\cite{cf,ex,ms}. With this in mind we develop a unified framework to 
combine the oscillation new physics with, presumably related,
direct-interaction new physics. We apply the general formalism to a number of
cases with an eye to accommodating all of the positive
signals for new phenomena within a three-family framework. 
We focus mainly on the picture where the favored\cite{hl} 
$\nu_e\leftrightarrow \nu_\mu$  MSW\cite{wm} oscillation is the solution
to the solar  neutrino observations, a $\nu_\mu \leftrightarrow\nu_\tau$
oscillation is the solution to the  atmospheric neutrino anomaly\cite{tg}  
and a direct interaction of the type $\nu_\mu + n \rightarrow e^- + p$  at
the detector and/or $\pi^+ \rightarrow \mu^+ + \nu_e$  or $\mu^+
\rightarrow e^+ + \bar\nu_e +  \nu_\mu$ at the source contribute
substantially to  the LSND signal\cite{Lc,nx} for the ``$\nu_e$  
appearance"  at 
the few tenths of a percent level.  The direct interaction effects are 
flight-path independent and, moreover, they 
are clearly too small to explain the  large solar and atmospheric anomalies
that have been reported.  At the  same time, the strictly two--flavor 
$\nu_e \leftrightarrow \nu_\mu$  oscillations with currently preferred[9] 
mixing  and mass values,
are negligible in the short--baseline experiments of the  LSND type. To
complement the ``two two--flavor", $\nu_e\leftrightarrow \nu_\mu$ plus
$\nu_\mu\leftrightarrow \nu_\tau$, picture just
mentioned, we include a brief discussion of a simple ``one-scale" version of
three--flavor mixing within our framework\cite{cf}. We show the addition of
modest direct interaction effects extends the allowed
parameter space in an interesting way.

A clean and convenient parameterization of the situation can be  achieved
by separating the problem into the production (source),  propagation (with
or without matter enhancement effects)  and  observation (detector)
components\cite{gkl,gk,yg}.  First we introduce a  working
phenomenological model which includes lepton-flavor violating, four-Fermi 
 interactions of purely leptonic and of semileptonic
types.\footnote{The questions of the detailed form and origin of the
neutrino mass matrix are not addressed
here.} Next we present the setup to describe the production,  propagation
and detection and define the ``appearance probability" for the situation
where the neutrinos are relativistic, the source is a muon or  charged pion,
the propagation is in vacuum and the detector is a neutron or proton (or
nucleus) target.
We then survey serveral patterns of dominant interactions
suggested by the data.  We compare  these patterns with those presented by the
minimal left-right symmetric model\cite{mrgs,ndg} and the minimal
supersymmetric model  augmented by a minimal set of discrete R-parity violating
terms\cite{sw,hdrev}.  Considering the LSND\cite{Lc,nx} result by itself, we
make a study  of the shift in allowed mass and mixing angle parameter
space that is  afforded by including direct interactions. In our 
concluding discussion, we point out several open possibilities.

\section{A Working Model}
To treat the bulk of the current accelerator
data and  to illustrate the general ideas, it suffices to take a model with
two lepton families and one quark family with charged current,  
effective four-Fermi interactions.  The model semileptonic (S) and
leptonic (L)   effective Lagrangian takes the form  $L_{eff} = L^S + L^L$,    
where\footnote{The factor of ${1\over 2}$ in the definitions 
$P_L={1\over 2} (1-\gamma_5)$ and $P_R ={1\over 2} (1+\gamma_5)$ leads to the
factor $2^{3/2}$ 
when combined with the conventional $G_F\over \sqrt{2}$ factor in the
four-Fermi interactions. Neutral current SM terms are also contained the $L^L$
term by Fierz symmetry of $(V-A)\times (V-A)$.}
\begin{equation}
L^S = 2^{3/2} G_F O_A (K^h_{Aij} \bar l_i \Gamma_A  P_h U_{ja} \nu_a 
)  + h.c. \label{eq:LS}
\end{equation}
\begin{equation}
L^L = 2^{3/2} G_F F^h_{Aijkm} \bar l_i \Gamma_A P_h U_{ja} \nu_a (\bar l_k
\Gamma_A P_h U_{mb} \nu_b)^\dagger, \label{eq:LL}
\end{equation}
where the coefficients K and F give the amount of admixture of $h=L,R$ projection,
the type of Dirac matrix structure and the lepton flavors contained in
$L_{eff}$. Possible crossed terms between $L$ and $R$ give contributions to the
unpolarized cross  sections that are proportional to $m_\nu$'s and we ignore
these effects here. The hadronic operators are of form 
$O_A = \bar d \Gamma_A( aP_L+bP_R) u$. All repeated indices
are summed: $A = S, V, T$ (matching  Dirac gamma matrices ${\bf 1},
\gamma$ and $\sigma$), $h=L,R,$ lepton flavor indices $i, j, k, m$ and  mass
eigenstate 
indices $a, b$.  The $U_{ja}$  are the unitary transformation  matrices
between the flavor states and the mass eigenstates.  

In a two--flavor mixing scheme where $\nu_e$ and
$\nu_\mu$ stand for the fields whose $P_L$ projections couple  with $e$
and $\mu$  to the W-boson, we express $\nu_e$ and $\nu_\mu$
in  terms of the ``mass eigenfields" $\nu_1$ and $\nu_2$, and  write $\nu_e =
cos \theta\; \nu_1 + sin\theta\; \nu_2$   and  $\nu_\mu= -sin\theta\; \nu_1 +
cos\theta\; \nu_2$.   Several examples of the coefficients $K$
and $F$  are tabulated for the $SM$ and two familiar extensions in
Table ~(\ref{tab:coef}).  In  the ultrarelativistic limit, an approximate 
Fock space of
flavor states can be constructed\cite{gkl,gk}, while the $|\nu_1>$ and 
$|\nu_2>$
states propagate with  the simple plane wave factors $exp\; i({\bf x}\cdot
{\bf p}- E_{1,2} t)$, leading at order $(\Delta  m^2)/E$ to the usual mass,
energy and distance dependent argument of  the oscillation probability.
The task at hand is to fold this  oscillation effect with the source and
detector lepton--flavor violating  effects to produce an ``appearance
probability" in terms of a convenient,  uniform parameterization.

\begin{table}
\caption{Some relevant expressions for $K$ and $F$ coefficients
in the SM, $L-R$ and $MSSM + \slash R$ models. The $L-R$ and $\slash R$ 
entries should be divided by $G_F 2^{3/2}$. 
} \label{tab:coef}
\begin{tabular} {|c|c|c|c|c|c|}    \hline
 & $K^L_{Vii}$ & $K^L_{V12}$ & $K^L_{S12}$ & $F^L_{V1122}$ & $F_{V211m}$ \\
\hline
$SM$ & 1 & 0 & 0 & 1 & 0 \\ \hline
$L-R$ & 0 & 0 & 0 & ${\bar h_{21}\bar h^*_{12}\over M^2_{\Delta_L}}$ & 
${\bar h_{11}\bar h^*_{2m}\over M^2_{\Delta_L}}$ \\ \hline
$\slash R$ & $\sum\limits_k {|\lambda'_{iik}|^2\over M^2_{\tilde d^k_R}}$ & 
$\sum\limits_k {\lambda^{'*}_{21k}\lambda'_{11k}\over M^2_{\tilde d^k_R}}$ & 
${\lambda^{'*}_{311}\lambda_{231}\over M^2_{\tilde e_3}}$ & 
${-|\lambda_{123}|^2\over 2M^2_{\tilde e^3_R}}$ & 0 \\ \hline
\end{tabular}
\end{table}

\section{Analysis of an Accelerator ``Appearance" Process}
Consider $\pi^+ \rightarrow
\nu + \mu^+$  decay to be the  source of neutrinos, where the
transition matrix element to a mass eigenstate $|\nu_b>$  reads $M_b^S =
<\nu_b, \mu^+|L^S |\pi^+>$, with $L^S$ given in Eq. ~(\ref{eq:LS}).   
Consider the reaction $\nu  +  N_i(A,Z-1) \rightarrow N_f(A,Z) + e^-$ to
be the detection  mechanism, with matrix element $M_b^D = <e^-,N_f| L^S
|N_i,\nu_b>$.  We may now write the  $\mu\rightarrow e$ transition
probability  in the ultrarelativistic neutrino limit as\cite{gkl,gk}
$P_{\mu\rightarrow e}  \sim 
|\Sigma_b M_b^D e^{-iE_bt} M_b^S |^2 ,
$
 where the sum is over mass eigenstates and the equality holds to leading
order in $m_\nu /E$.  Spelling out this probability in the notation of Eq.
~(\ref{eq:LS}), we have the expression
\begin{equation}
\begin{array}{lll}
P_{\mu\rightarrow e} & \sim & (2^{3/2} G_F)^4| \Sigma_b
(K^h_{A2j})^*
K^h_{B1k}(\bar \mu \Gamma_A P_hU_{jb} \nu_b)^*\times\nonumber \\
& & \bar e \Gamma_BP_hU_{kb} \nu_b  e^{-iE_bt}<0|O_A|\pi^+> <N_f|O_B|Ni>|^2.
\label{eq:PS} \\
\end{array}
\nonumber
\end{equation}
The sums over repeated indices $A, B, h, j, k$ and a sum over spins  are
implicit.  
In the same fashion as above, the
$\mu^+ \rightarrow  e^+ + \nu  + \bar \nu$  process can be considered to 
be the
source of neutrinos, with the effective Lagrangian $L^L$ replacing $L^S$ in
the source matrix element. The ``appearance" sought is then
$\bar \nu + N_i(A,Z+1)\rightarrow N_f (A,Z) + e^+$.  The counterpart to 
Eq. ~(\ref{eq:PS}) reads: 
\begin{equation}
\begin{array}{lll}
P_{\mu\rightarrow e} & \sim & (2^{3/2} G_F)^4 \Sigma_c|\Sigma_d
(K^h_{B1k})^*
F^h_{A2j1m} (\bar \mu \Gamma_A P_hU_{jd} \nu_d) \times \nonumber \\
& & (\bar e \Gamma_AP_hU_{mc} \nu_c)^* e^{-iE_dt}
(\bar e \Gamma_BP_hU_{kd}\nu_d)^* <N_f|O_B|Ni>|^2. \\
\label{eq:PL}      
\end{array}
\end{equation} The expression in Eq. ~(\ref{eq:PS}) is appropriate to the 
LSND decay-in-flight (DIF) study
and Eq. ~(\ref{eq:PL}) to their decay-at-rest (DAR) study with $N_i=p$ and 
$N_f=n$.
To make the notation more concrete, let us illustrate it with several
examples.

\section{Purely Leptonic New Physics}
Our first example is applicable to the DAR situation, where $\mu^+$ decay
provides the $\bar \nu$'s at the detector. We assume only the lepton source
contains new, lepton--number violating interactions, and that they are
represented in the effective Lagrangian, Eq. ~(\ref{eq:LL}), by $(V-A)\times 
(V-A)$ terms.
The lepton--flavor violating case of interest to us has $i=2,\; j=1,\; k=1$ and
$m=1,2$ or 3. Applying Eq. ~(\ref{eq:PL}) to this situation, one reads off, 
for $\mu^+\rightarrow e^+\nu\bar\nu$ and $ \bar\nu_e + N(Z)\rightarrow
e^+ + N(Z-1)$,
\begin{equation}
\begin{array}{lll}
P_{\mu\rightarrow e} & \sim & \{ ( 2^{3/2} G_F)^4|\bar e\gamma_\lambda
P_L \nu < N_f| O^\lambda_V | N_i>|^2 |K^L_{V11} |^2|\bar\mu\gamma_\sigma 
P_L\nu \bar \nu \gamma_\sigma P_Le|^2\} \nonumber \\
& &\times\sum\limits_c \mid\sum\limits_d  F^L_{V2211} U^*_{2d} e^{-iE_d t} 
U_{1d} U^*_{1c}+  F^L_{V211m} U^*_{1d}e^{-iE_d t} U_{1d} U^*_{mc}|^2 . 
\label{eq:PLs} \\ 
\end{array}
\end{equation}
With the new physics confined to the lepton--number violation in the $\mu^+$
decay as described, we have $K^L_{V11}=1=F^L_{V2211}$ and
$F^L_{V211m}\not=0$. The factor in braces is exactly the one appropriate 
to the unsupressed sequence $\mu^+\rightarrow e^+ \bar\nu_\mu \nu_e$ and 
$\bar\nu_e+N(Z)\rightarrow N(Z-1)+e^+$.
The $P_{\mu\rightarrow e}$ probability can now be read off from Eq. 
~(\ref{eq:PLs}), and for two flavor mixing we obtain
\begin{equation}
\begin{array}{lll}
P_{\mu\rightarrow e} & = & \sum\limits_{m=e,\mu,\tau} [
tan^2\psi_m + (1- tan^2\psi_m)  sin^22\theta sin^2 x] \\
& + & [\tan\psi_e sin 4\theta  sin^2x\nonumber \\
& + & 4 tan \psi_e sin 2\theta sin \varphi_e sin x(cos\varphi_e cos x - 
cos 2 \theta sin \varphi_e sin x)]  ,\nonumber \\
\end{array}
\end{equation}
where $2x=(E_2-E_1)t$ and 
where $F^L_{V211m}\equiv tan\psi_m e^{2i\varphi_m}$ has been defined so that
a uniform 
``all angles" representation of the appearance probability could be achieved. 
The limit 
$\varphi_m \rightarrow 0$, keeping only an $m=\mu$ value and keeping 
leading powers of $\psi_m  <<1$,
leads to the formula of Ref. \cite{yg} for the case that the direct 
lepton--flavor violation is only in the $\mu$-decay source interaction.

In Fig. ~(\ref{fig:DAR}) we show the $P_{\mu\rightarrow e}=0.003\pm 
0.001$\cite{Lc} contours in the $sin^22\theta$, $\Delta M^2_{\mu e}$ 
parameter
plane for $\psi_e\equiv F^L_{V 2111}$ values of 0 and 0.04, where
$\psi_\mu=\psi_\tau=\varphi_m=0$ is assumed.\footnote{Since we are showing only
trends and not fits, we adopt the Gaussian ``toy model" used for illustrating
oscillations in Ref. \cite{pdg} with $b_0$ and $\sigma_b$ appropriate to 
LSND.} The
minimal $L-R$ model\cite{mrgs} is relevant to the case under consideration, 
since it 
generates lepton--flavor violating, purely leptonic, four-Fermi,
$(V-A)\times (V-A)$ interactions by exchange of an iso--triplet Higgs particle.
There is no $(V-A)\times (V-A)$ $\mu^{+}\rightarrow 
e^{+}+\bar \nu_{e}+\nu_{x}$ effect in 
the $MSSM + \slash R$ model, contrary to the assertion in Ref. [14].
In the notation of Eq. ~(\ref{eq:LS}) we have $2^{2/3} G_F F^L_{Vijkm} =
{\bar h_{jk}\bar h^*_{im}\over M^2_{\Delta_L}}$ as the identification of the
relevant coefficients in the effective Lagrangian, Eq. ~(\ref{eq:LL}). 
Yukawa coupling
matrices are designated by $\bar h_{ij}={1\over 2}(h_{ij}+h_{ji})$ and
$M_{\Delta_L}$ is the mass of the singly-charged member of the Higgs triplet,
following the notation of Ref. \cite{ndg}. The ranges of allowed values of 
${\bar h_{11}\bar 
h_{2m}\over M^2_{\Delta_L}}$ for the $L-R$ model are shown in Table 
~(\ref{tab:bounds}). 
{\em It is surprising that a considerable flexibility in choosing relevant 
$L-R$ model parameters is still allowed by all experimental
constraints\/}. 

\begin{figure}
\caption{The LSND 1$\sigma$ allowed region ($P=0.002 - 
0.004$ for DAR) for no direct interaction new physics (solid curves) and 
purely leptonic new physics strength $\psi=0.04$ (dashed curves). Nonzero 
new physics phase makes the curves slightly lower for 
$\Delta M^{2} < 1.8 eV^{2}$. Also included are limits from Bugey 
(dash-dotted curve) and E776 (dotted curve).} \label{fig:DAR}
\end{figure}

\begin{table}
\caption{Relevant new physics couplings and upper bounds.
Sources of the bounds are {\em other\/} than neutrino oscillation 
experiments.} \label{tab:bounds}
\begin{tabular} {|l|c|c| } \hline
{\bf Coupling} & {\bf Bound} & {\bf Source} \\ \hline
$|\bar h _{11}\bar h^*_{21}|/(M_{\Delta_L^2} G_F 2\sqrt{2})$ & $7.6 \times
10^{-2}$ & $\nu e,\nu_\mu e, \bar\nu e$ \\ \hline
$|\bar h _{11}||\bar h^*_{22}|/(M_{\Delta_L^2} G_F 2\sqrt{2})$ & $1.1 \times
10^{-1}$ & $\mu\rightarrow e \nu\bar \nu$ \\ \hline
$|\bar h _{11}||\bar h^*_{23}|/(M_{\Delta_L^2} G_F 2\sqrt{2})$ & $9.1 \times
10^{-1}$ & $(g-2)_\mu$ \\ \hline
$|\sum\limits_k \lambda^{'*}_{21k} \lambda_{11k}/(M^2_{\tilde d_R^k} G_F
2\sqrt{2})|$ & $2.5\times 10^{-7}$ & $\mu Ti\rightarrow e T i$ \\ \hline
$|\lambda^{'*}_{211}||\lambda_{231}|/(M^2_{\tilde e_R^k} G_F 2\sqrt{2})$ & $3.3
\times 10^{-2}$ & $\tau\rightarrow \ell \nu\bar \nu$ and $\pi\nu$ \\ \hline
\end{tabular}
\end{table}

\section{The Same New Physics at Source and Detector} 
A case that injects the same new semileptonic physics at source and 
detector in the
DIF situation is exemplified by the pure $(V-A)\times (V-A)$ choices
$K^L_{V22}=K_{V11}^L=1$, which follow from the standard model, and
$K^{*L}_{V21} = K^L_{V12}\equiv tan \psi e^{2i\varphi}$ which defines
the new physics.\footnote{We use the same symbols $\psi$ and $\varphi$ to
parameterize the different cases for economy of notation. The context makes
clear that these are independent parameters for the different cases.}
 The $P_{e\rightarrow \mu}$ expression that results
when two-flavor mixing is assumed reads
\begin{equation}
P_{\mu\rightarrow e}(DIF)  =  4 tan^2\psi\; cos^2x + sin^2 2\theta 
sin^2 x 
 -  2 tan \psi sin 2 \theta sin 2 \varphi sin 2 x \label{eq:PSsd}
\end{equation}
\noindent for DIF.\footnote{On this point, we completely disagree with Ref.
\cite{yg}, which asserts that there is no effect when the physics at 
source and detector are the same.}  On the other hand,
\begin{equation}
\begin{array}{lll}
P_{\mu\rightarrow e}(DAR) & = &  tan^2\psi + sin^2 2\theta 
sin^2 x\; (1 - tan^2\psi) \nonumber \\
& + & tan \psi sin 2 \theta\; [- sin 2 \varphi sin 2x + 2 cos
2 \varphi cos 2 \theta sin^2x] \label{eq:PLd} \\
\end{array}
\end{equation}
\noindent 
for DAR, where only the detection is affected by the new interaction.  In
writing Eq. ~(\ref{eq:PLd}), we have assumed that there is no lepton--flavor 
violating new
physics interaction in $\mu$-decay. If $sin^2x << sin^2\psi$, then 
$P_{\mu\rightarrow e} (DIF)\simeq
4 P_{\mu\rightarrow e}(DAR)$, so a {\em combination\/} of $L^S$ and $L^L$ 
effects
would have to be invoked to produce the $P_{\mu\rightarrow e} (DIF) \simeq
P_{\mu\rightarrow e}(DAR)
\simeq 0.003$ result reported by the two
LSND experiments. Fig. ~(\ref{fig:DIF}) shows plots of the $P_{\mu\rightarrow
e}(DIF)=0.0026\pm 0.0011$ boundaries in the $\Delta M^2_{\mu 
e}\leftrightarrow sin^2 2\theta$ parameter plane for several choices of
$\psi$ with $\varphi=0$. For smaller $\psi$ values, $\psi\ltsim 0.02$, the
modifications to the $\psi=0$ curve have the same general character as those of
Fig. ~(\ref{fig:DAR}), moving the curves leftward and downward. A dramatic 
change occurs when
$\psi \gtsim 0.03$, however. In Eq. ~(\ref{eq:PSsd}) the large term $4 
tan^2\psi cos^2x$ tends
to ``overshoot" the input value of $P_{\mu\rightarrow e}$ as $tan^2\psi$ grows,
so $x$ needs to increase to compensate this growth. But the second term
grows as $x$ grows, so $sin^2 2\theta$ must control this growth and therefore a
{\it maximum} $sin^2 2 \theta$ value is established for larger $tan^2\psi$
values. It is the factor four in the first term of Eq.
~(\ref{eq:PSsd}), reflecting the compound effects of source and detector, 
that produces the new effect at rather modest $tan^2\psi$ values.

\begin{figure}
\caption{The LSND 1$\sigma$ allowed region ($P=0.0015 - 
0.0037$ for DIF) for no new direct interaction physics (solid curves) and 
for $P=0.0026$ with $\psi=0.01$ (dotted curve), $\psi=0.02$ (dot-dashed 
curve) and $\psi=0.03$ (dashed curve) when there is the same new physics 
at the source and detector.} \label{fig:DIF}
\end{figure}

To make the connection to the minimal $\slash R$ SUSY model,\cite{hdrev} we 
read off $K^{*L}_{V21}=K_{V12}= \sum\limits_k
{\lambda^{'*}_{21k}\lambda'_{11k}\over M^2_{\tilde d^k_R}}\cdot {1\over
2^{3/2}G_F}$ from the semileptonic effective Lagrangian that follows
from exchange of the $\tilde d_R^k$ squark, with $k$ the
family index, in the notation of Ref. \cite{vbg}. In this model, the neutral
current $e\leftrightarrow \mu$ transition operator has the same
coefficient, and the bound from $\mu Ti\slash\rightarrow eTi$ 
\cite{kkl,pr} is
severe, as listed in Table ~(\ref{tab:bounds}), making this case academic in 
the minimal $\slash R$ SUSY model. At the moment we cannot offer a
well--motivated model with the structure $K^{*L}_{V21}=K^L_{V12}$, but we
present the above to illustrate the structure of the case where the
source and detector new physics is the same.

\section{New Physics Only at the Detector} 
Because the DAR and
DIF experiments of the LSND collaboration have different backgrounds
and different systematic errors, the reported agreement between the results of
the two experiments suggests a common origin\cite{nx}. We
point out here that a lepton--number violating effective interaction
that is only operative in the detection processes tends to produce this 
effect.

Consider the effective Lagrangian case with
$K^L_{V22}=K^L_{L11} = 1$ for the standard model, $K^L_{V12}\not= 0=
tan \psi e^{2i\varphi}$ and all other coefficients negligible 
for the new physics. We find that, in the two--neutrino mixing case, the same
parameterization, Eq. ~(\ref{eq:PLd}), applies to 
both the DAR and DIF cases {\em even with the direct, lepton--number 
violating
interactions effects combined with those of the oscillations\/}. In 
principle the agreement between the DAR and DIF 
measured $\nu_\mu\leftrightarrow \nu_e$ oscillation probabilities could be
represented by $tan^2\psi$, a purely direct interaction effect with
$tan^2\psi\simeq 0.003$

In surveying our example models, the minimal $SU_L(2)\times SU_R(2)\times
U(1)_{B-L}$
weak interaction model and the MSSM without
$R$-parity, one finds that the above parameter choice
$K^L_{V12\not=0}$ and all other new physics $K^L_{Aij}=0$, does not
occur. The instructive case $K^L_{S12}\not= 0$, $K^L_{S21}=0$ and with 
$K^L_{Vij}=0$ by choice, {\em does\/} arise from the $R$--parity violating
interactions, however, and we summarize this case next.

\section{Different New Physics Lorentz Structures}
Up to this point, we have considered only cases where the new
interaction, charged current effective four--Fermi Lagrangian has the
same $(V-A)\times (V-A)$ structure as the standard model. This ensured
all matrix elements of interest had the same structure and the
dynamics of the source and detector physics could be lumped into an
overall factor that is normalized away in the definition of the lepton
``flavor change", or ``oscillation" probability. If the Lorentz
structure of the charged current $\times$ charged current interaction is {\it 
not} the same as that of the standard model, then the analysis of
$P_{\mu\rightarrow e}$ necessarily brings in details of the matrix elements
at the source and/or detector. 

In the situation mentioned above where
$K^L_{S12}\not= 0$ is the only relevant new physics, and the scalar
interaction is of the form $(S-P)_{lepton}\times (S+P)_{quark}$, 
the matrix elements of the new physics and those of the standard model
operators are distinctly different. Dividing out a common factor from the
source, the DAR case looks schematically like, in the two--flavor mixing
case,
\begin{equation}
\begin{array}{lll}
P_{\mu\rightarrow e} & \sim & |\bar e\gamma_\lambda P_L\nu <N_f |
O^\lambda|N_i>(-sin 2\theta sin x e^{ix}) \nonumber \\
& + & K^L_{S12}\bar e P_L\nu < N_f| O_S|N_i>(1-2cos^2\theta sin x
e^{ix}) |^2. \\
\end{array}
\end{equation}
\noindent The new physics, lepton--number violating interaction effects depend
both on the parameters $(K^L_{S12}, \theta$ and $x$) and on
the comparative role of the matrix elements: $\bar e\gamma_\lambda
P_L\nu <N_f|O^\lambda|N_i>$ in the standard model term and $\bar e
P_L\nu <N_f| O_S|N_i>$ in the new interaction. With $O_S=\bar u
P_R d$, as in the $R$--parity violating model where
$K^L_{S12}=\lambda^{'*}_{311}\lambda_{231}/(M^2_{\tilde e_3} G_F
2\sqrt{2}), K^L_{S21}=0$ and $K^L_{V12}=K^{*L}_{V21}<<K^L_{S12}$, we can
evaluate the spin--averaged transition probability for a $\nu + n
\rightarrow e + p$ detector transition to illustrate the DAR application.
Simplifying the $\beta$-decay matrix element to pure $V-A$ for
illustration purposes, we find
\begin{equation}
P_{\mu\rightarrow e}\sim\sin^2 2 \theta sin^2 x +|K^L_{S12}|^2
{t(t-2M^2_N)\over 4(s-M_N^2)^2} (1-sin^2 2\theta sin^2 x), \label{eq:PSnl}
\end{equation}
where $M_N$ is the mass of the nucleon,
$s= (P_\nu+P_n)^2$ and $t=(P_\nu-P_e)^2.$ In writing Eq. 
~(\ref{eq:PSnl}) we have
dropped the cross term, which is proportional to $m_e$, and we have not
displayed a final state phase space integration. The matrix element
effects are clearly crucial, as is shown by the suppression of the
cross term and the nontrivial nature of the relative kinematical
factor between the $(V-A)\times (V-A)$ standard model structure and the
$(S-P)\times (S+P)$ new physics structure. For energies appropriate to
LSND, the kinematical factor ratio that multiplies the new physics term
is small (of order $|t/m^2_N|\sim (30 MeV/1 GeV)^2)$, so it is unlikely
that such terms can contribute significantly to a $P_{\mu\rightarrow
e}$ signal. For completeness, we mention that the MSSM with $\slash R$, where 
$|\lambda^{'*}_{311}\lambda_{231}|/M^2_{\tilde e_3}G_F 2\sqrt{2}\leq 2
\times 10^{-4}$\cite{pr}, requires a {\em kinematical enhancement\/}
to bring it up to the order of $10^{-3}$ relevant to the LSND reported
result. The important lesson to stress, however, is that one must look
at the detailed matrix element evaluations to draw conclusions in these
cases where the Lorentz structure of the effective Lagrangian's new physics
terms is different from that of the standard model.

\section{Three--Family Mixing Application}
Our illustrations so far have relied on two--family mixing
parameterizations, which are appropriate to the $M_3 >> M_2 >> M_1$
type of hierarchy of neutrino masses. The framework we present
applies to any number of families, so we discuss a simple, ``almost
viable" version of three--family mixing as our last illustration
\cite{cf}. The starting, simplifying assumption is that
$M_1\sim M_2<<M_3$, which reduces the three--family mixing to a
``one-scale" problem; namely $\Delta M^2_{13}\simeq \Delta M^2_{23}\equiv
\Delta M^2$ with $\Delta M_{12}^2\simeq 0$. Returning to the DAR case
with new physics at the source and a restriction to $m=1$ in the flavor
sum to reduce clutter in the notation, we have
\begin{equation}
P_{e_\mu}=|-2iU^*_{23}U_{13}sin xe^{-ix}+tan \psi e^{2i\varphi}(1 -2i
|U_{13}|^2 sin x e^{-ix})|^2, \label{eq:PL3}
\end{equation}
where $F_{V2111}\equiv tan \psi e^{2i\varphi}$. With the parameterization
$U_{23}=sin\theta_{23}cos 
\theta_{13}$ and $U_{13}= sin \theta_{13}e^{-i\delta_{13}}$\cite{pdg}, one 
can write out an expression similar to, but even more 
obscure than Eq. ~(\ref{eq:PLd}). The form shown in Eq. ~(\ref{eq:PL3}) 
makes the point 
that the first term alone gives the result of Ref. \cite{cf}, and the new
interactions produce a leading $tan^2\psi$ term plus other terms similar to
those in Eq. ~(\ref{eq:PLd}). The flexibility afforded by the inclusion 
of the
direct interaction is shown in Fig. ~(\ref{fig:3-1}), where the effect of 
increasing $\psi$ on
the $tan^2\theta_{13}$ vs. $tan^2\theta_{23}$ plot is shown to be significant
for $\psi\simeq 0.04$.

\begin{figure}
\caption{For three-flavor, one-scale mixing, the LSND 
1$\sigma$ allowed region (DAR) for no new direct interaction physics 
(solid curves) and purely leptonic new physics strength $\psi=0.02$ 
(dotted curves) and $\psi=0.04$ (dashed curves).} \label{fig:3-1}
\end{figure}

\section{Discussion}
We have developed and applied a compact analysis that treats neutrino
flavor--changing phenomena -- oscillations plus direct interactions -- in a
unified manner. We showed standard oscillation parameter plots can be
shifted significantly for rather modest values of $\psi$, an angle
parameterizing direct interactions.
We also showed the positive DAR result reported by LSND\cite{Lc} can be
accounted for by direct interaction effects in the $L-R$ model.

Open possibilities for further applications include the unified treatment of
the neutrino masses and direct interactions within a model like the 
$MSSM + \slash R$ model and the applications to general three--family and 
three--family plus sterile neutrino pictures.

Whether direct lepton--flavor changing effects turn out to be important or not,
the framework offered here should provide an efficient means to survey the
possibilities.

\section{Acknowledgements} 
We thank Kerry Whisnant and Herman Munczek
for discussions and Jack Gunion and Tao Han for a timely $\slash R$ 
suggestion. L.M.J. acknowledges support from the University of Kansas 
Dissertation Year program. This work was supported in part by U.S. DOE Grant 
No. DE-FG02-85ER40214.

{}

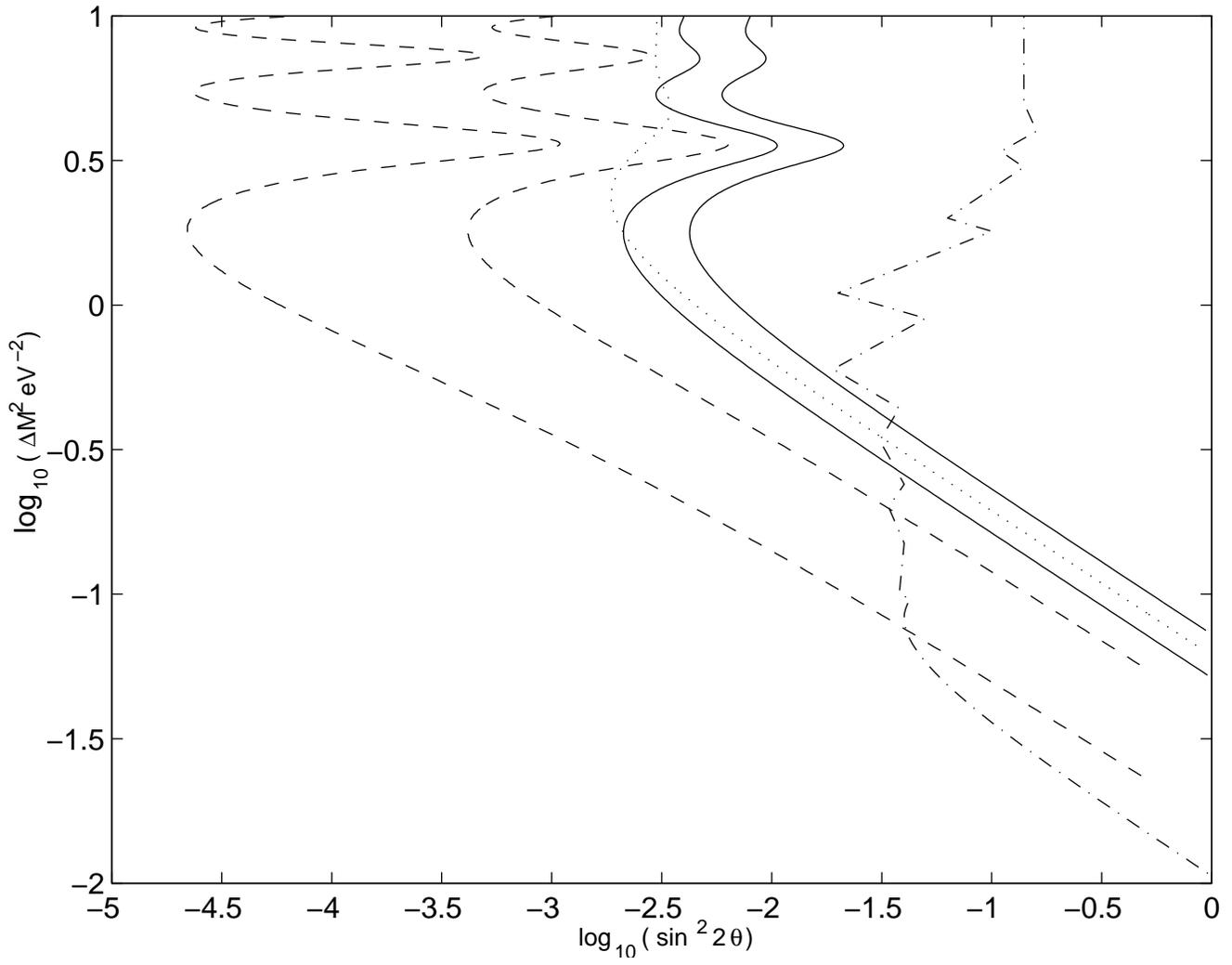
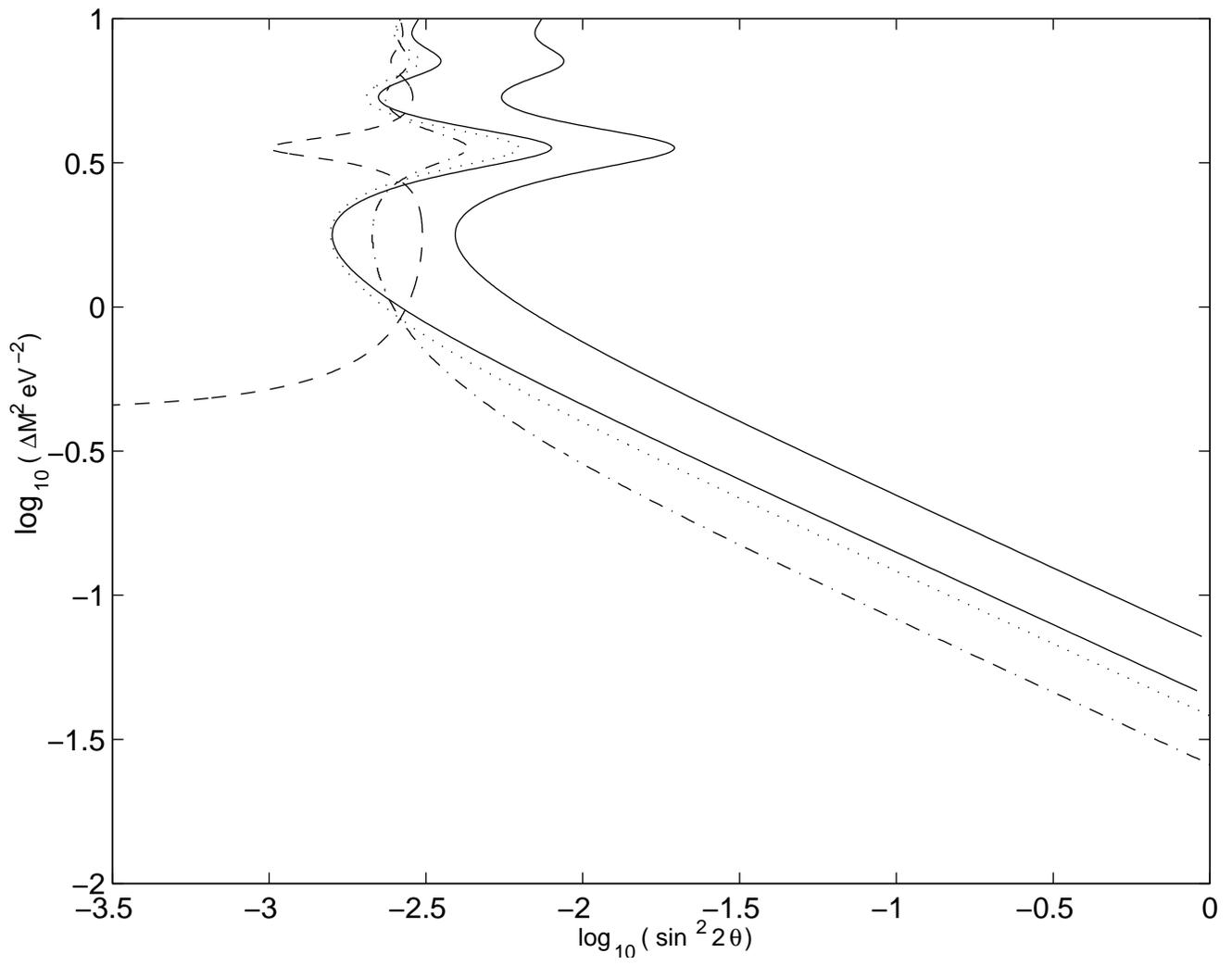
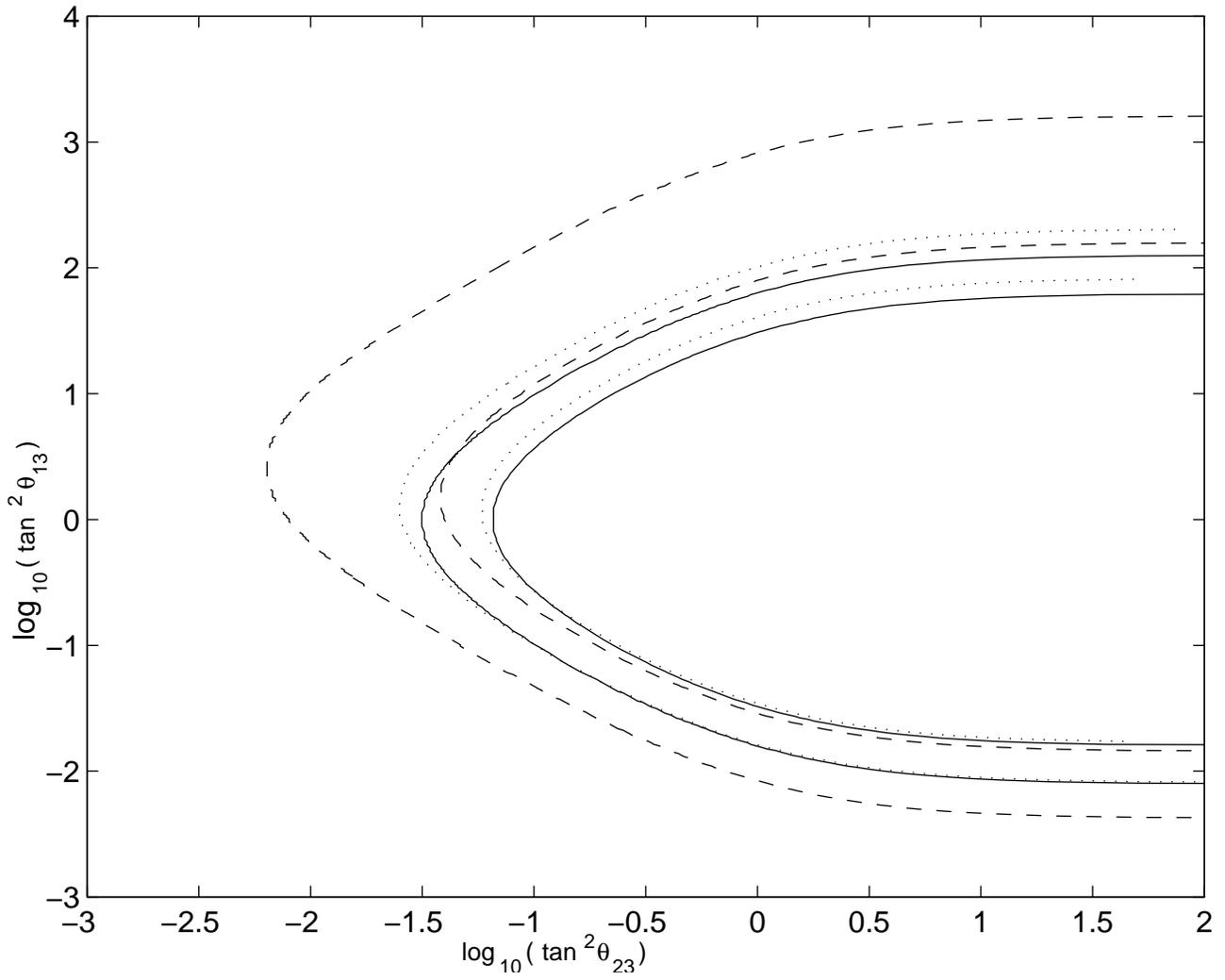
\end{document}